\documentclass[aps,prb,superscriptaddress,showpacs,twocolumn,amssymb]{revtex4}
\usepackage{graphicx}

\begin{document}
\setcounter{page}{1}

\title
{Statistical work-energy theorems in deterministic dynamics}

\author{Chang Sub Kim}

\email{cskim@jnu.ac.kr}

\affiliation {Department of Physics,
Chonnam National University,
Gwangju 500-757, Republic of Korea}


\begin{abstract}
We theoretically explore the Bochkov-Kuzovlev-Jarzynski-Crooks work theorems in a finite system subject to external control, which is coupled to a heat reservoir.
We first elaborate the mechanical energy-balance between the system and the surrounding reservoir and proceed to formulate the statistical counterpart under the general nonequilibrium conditions.
Consequently, a consistency condition is derived, underpinning the nonequilibrium equalities, both in the framework of the system-centric and nonautonomous Hamiltonian pictures and its utility is examined in a few examples.
Also, we elucidate that the symmetric fluctuation associated with forward and backward manipulation of the nonequilibrium work is contingent on time-reversal invariance of the underlying mesoscopic dynamics.
\end{abstract}


\maketitle

\section{Introduction}
\label{introduction}

Nonequilibrium (NEQ) transformation is not characterized by a simple thermodynamic relation and even relevant variables to be specified are not given, in general, a priori.
The Jarzynski identity rarely provides researchers with a useful avenue to extract a definite, equilibrium information of thermodynamic systems driven away from equilibrium \cite{JarPRL1997,JarPRE1997}.
The Crooks relation follows to describe the symmetric work-fluctuation in statistical dynamics of the systems \cite{Crooks1998,Crooks1999}.
The two, widely known as the NEQ work theorems are found to be intimately related to the earlier study by Bochkov and Kuzovlev \cite{Bochkov1977}.
Aside from their functionality \cite{Ritort2005}, the NEQ work theorems deliver an insight into the nature of the second law of thermodynamics in the physical regime where indeterminacy matters beyond conventional bulk thermodynamics in the context of the disparate fluctuation theorems (FTs) \cite{Evans1993,Evans1994,Gallavotti1PRL995,Gallavotti1JST995,Kurchan1998,Lebowitz1999,Crooks2000}.

The NEQ work theorems were framed for a thermodynamic system on which one cannot perform a work precisely as instructed by a predetermined protocol \cite{Jar2008}.
For instance, in single-molecule stretching experiments, one inevitably ends up with
a stochastic signal in the measured, force versus extension data \cite{Strick2001}.
The free energy (FE) difference between two equilibrium states, supposedly undergone
an irreversible transformation, is given by an \textit{equality}, called the Jarzynski equality (JE).
A dedicated feature is that there appears only a single temperature in the JE, albeit the system is perturbed from initial equilibrium \cite{Cohen2004,Jarzynski2004,Cohen2005,Baule2006}.
Over the years, a great deal of research effort has been devoted to establish the NEQ formulation more rigorously using the various mathematical methods in the theoretical side and to test the idea in the real and computer experiments \cite{Hummer2001,Hatano2001,Jensen2002,Liphardt2002,Onoa2004,Collin2005,Imparato2005,Harris2007,Seifert2008,Harris2009,
Chen2010,Hummer2010,Palassini2011,Jar2011,Gupta2011,Echeverria2012,Jar2013PRE,Dhakal2013,Ritort2014}.
Lately, research effort has broken a new ground to explore more intricate transition between kinetic states of systems \cite{Alemany2012}.

In this paper we consider a finite system, maintained in a heat reservoir, under controlled manipulation by an external agent.
The statistical correlation of the system with the surrounding reservoir gives rise to a momentum-dependent thermostatting mechanism, which was not taken into account in the original Hamiltonian derivation of the JE \cite{JarPRL1997,Jarzynski2004}.
The thermostatting damping force is embraced explicitly in our formulation, together with the time-dependent control force.
The control force is generically macroscopic because it describes coupling of an apparatus to its conjugate phase-observable of the system.
Therefore, the entailed NEQ dynamics is mixed or \textit{mesoscopic} on time scale in the sense that the coarse-grained damping force and the macroscopic external coupling conjoin to determine the microscopic time-evolution of the system.

The goal of our endeavor is to present a statistical-mechanical formulation of the NEQ work theorems
in the finite, dissipative systems governed by the proposed, mesoscopic dynamics.
The main concerns are to investigate the effect of the phase-volume contraction due to thermostatted dissipation and other nonpotential fields,
as a consequence to derive the consistency criterion that undergirds the NEQ work theorems,
and to attentively clarify the subtlety of the notion of thermodynamic work in the NEQ measurements.
We hope to add a further insight into our understanding of the fundamental relation between dissipation
and irreversibility.

The proposed, extended dynamics and corresponding generalized Liouville equation are partly based on the early formalism by McLennan, for stead-states driven by external nonconservative forces \cite{McLennan1959}.
The subsequent analysis utilizes a formal solution to the modified Liouville equation, describing the NEQ distribution at a co-moving point in phase space,
which is often called the Kawasaki representation \cite{EM2008}.
The Kawasaki method provides a closed expression for far-from equilibrium distributions and was previously applied to investigating the transient FT in thermostatted fluid systems \cite{ES2002}.
Employing the analytical representation for the NEQ ensemble density, we show that the NEQ work theorems must be tightened in a thermally open system by a certain consistency criterion.

The JE is formulated for a unidirectional transformation that its experimental or computational test may be intended in single, forward irreversible setups where the reverse work cannot be measured  \cite{Jensen2002,Harris2009,Palassini2011,Gupta2011,Echeverria2012}.
Whereas, the Crooks work-fluctuation theorem (CWFT) is furnished for bidirectional, forward and backward measurements of a single small system \cite{Liphardt2002,Collin2005}.
We show that the latter, symmetric work theorem applies to the systems whose dynamics is \textit{constrictively} invariant under time reversal.
The constrictive invariance revealed rests on the mesoscopic symmetry of momentum-dependent dissipation.
Lately, there is a growing interest in the role of momentum, an odd-parity dynamical variable under time inversion, associated with the entropy productions in driven stochastic dynamics \cite{Spinney2012,Lee2013}.
We also notice a report which treats the microscopic reversibility of a Hamiltonian system under an external perturbation, but unlike our investigation, without dissipative mechanisms \cite{Monnai2012}.

The manipulating forces of the NEQ work measurements are usually installed as time-dependent parameters in the Hamiltonian \cite{JarPRL1997,Imparato2005,Hummer2010,Jar2013PRE,Ritort2014}.
Such a \textit{nonautonomous} description undesirably bears a delicacy in consistently defining the physical energy.
This point was previously called into question by others \cite{Rubi2008} and elevated the active discussions about a proper definition of thermodynamic work among researchers \cite{Peliti2008,Chen2008,Adib2009,Zimanyi2009,Rubi2011}.
Recently, an experimental study carefully concerned how to correctly define and measure thermodynamic work in small systems, which the authors found as a pertaining issue in pulling experiments \cite{Ritort2014}.
In this paper, we explore both the nonautonomous picture and, what we term, the \textit{system-centric} picture in a unified frame.
In the latter picture the time-dependent forces responsible for external work are included in the equations of motion directly and the energy is defined via the bare Hamiltonian.
Consequently, we explicate the two different representations of work and illuminate which picture brings about the conventional thermodynamic description.

Our work exclusively concerns deterministic dynamics that stochastic dynamics is not in the scope of our investigation.
For the latter, we refer to a recent review which contains the latest issues in stochastic thermodynamics with a complete list of references \cite{Seifert2012}.
The parallel NEQ formulation in quantum dynamics, which is also beyond the scope of the present article, can be
found in the other review \cite{Campisi2011}.

This paper is organized as follows.
In Sec.~\ref{ExEqMotions} we coin the extended Hamilton equations of motion and the related, modified Liouville equation, suitable for describing NEQ work measurement in a thermally open system.
Then, the mechanical work-energy theorems are established in Sec.~\ref{mechanical work-energy} and the corresponding statistical formulation is followed in Sec.~\ref{JarzynskiEquality} delivering the NEQ equalities with the upholding condition.
In Sec.~\ref{CrooksTheorem} the bidirectional, work-fluctuation theorems are considered with discussing the underlying symmetry of the generalized Liouville dynamics.
In Sec.~\ref{Examples} a few dynamical systems are examined to manifest the utility of the consistency criterion.
Finally, a summary and a conclusion are provided in Sec.~\ref{Conclusion}.

\section{Extended Hamiltonian dynamics}
\label{ExEqMotions}
The mechanical influences on a physical system are usually described by three types of forces: the coordinate-dependent, momentum-dependent, and parametric (time-dependent) forces.
We recapitulate here how these forces may enter the extended Hamilton equations of motion which, in turn, constitute the generalized ensemble dynamics of the identically prepared systems in phase space.

Let us suppose that a finite system under external control is coupled to a heat reservoir.
The total Hamiltonian $H_T$ can be written, in principle, as
\begin{equation}\label{total}
H_T = H + H_R + \Phi
\end{equation}
where $H$ and $H_R$ are the Hamiltonian of the system and the surrounding reservoir, respectively, and $\Phi$ denotes the boundary interaction between them.
The bare Hamiltonian of the system is specified by the generalized coordinates $q$ and momenta $p$ of the constituents of the system,
\[ H = H(q,p). \]
The Hamiltonian of the reservoir depends on the generalized coordinates and momenta $X,Y$ of the reservoir degrees of freedom as
\[ H_R = H_R(X,Y).\]
We shall assume that the interaction $\Phi$ depends only on the coordinates of the system and the reservoir to be written as
\[ \Phi = \Phi(q,X).\]

The total system must evolves in time obeying Hamiltonian dynamics at the microscopic level.
Accordingly, the NEQ Gibbs ensemble is described by the Liouville equation in the full phase space,
\begin{equation}\label{tLiouville}
\frac{\partial P}{\partial t} + \{P, H_T\} = 0
\end{equation}
where $P$ is the ensemble density of the total system and $\{P, H_T\}$ is the Poisson bracket \cite{Landau}.
By encapsulating the unknown information on $\Phi$ as a statistical correlation $\varphi$ between the system and the reservoir, one may write down the total ensemble density in the form,
\[P = \rho\rho_R(1+\varphi)\]
where $\rho_R$ denotes the canonical equilibrium density of the reservoir at an absolute temperature $T$.
The ensemble density of the system $\rho$ is a marginal density reduced from the total density via
\[ \rho(q,p;t) = \int d\Omega P(q,p;X,Y;t),\]
where $d\Omega\equiv dXdY$, assuming proper normalization in the full phase space.

In practice, time-evolution of the ensemble density of the system alone is of concern, which can be obtained by integrating Eq.~(\ref{tLiouville}) over the reservoir variables.
After some manipulation, the result is given as \cite{McLennan1959}
\begin{equation}\label{sLiouville}
\frac{\partial \rho}{\partial t} + \{\rho, H\} = - \sum\frac{\partial}{\partial p_i} \left(\rho{\cal R}_i\right)
\end{equation}
where use has been made of the identity,
\begin{equation}\label{r-force}
{\cal R}_i = \int d\Omega\left(-\frac{\partial \Phi}{\partial q_i}\right)\rho_R(1+\varphi)
\end{equation}
and the explicit expression of the Poisson bracket,
\begin{equation}\label{Poisson}
\{\rho, H\} \equiv \sum \left( \frac{\partial\rho}{\partial q_i}\frac{\partial H}{\partial p_i} -
\frac{\partial H}{\partial q_i}\frac{\partial \rho}{\partial p_i} \right)
\end{equation}
with summation running over all degrees of freedom in the system.

When the system and the reservoir is statistically uncorrelated, i.e. $\varphi\equiv 0$, ${\cal R}_i$ becomes
only coordinate-dependent to give
\[{\cal R}_i \rightarrow -\frac{\partial}{\partial q_i} V_R(q)\]
where $V_R \equiv\int d\Omega \Phi\rho_R$ which may be absorbed into the system Hamiltonian,
\[
\tilde H \equiv H + V_R.
\]
Then, Eq.~(\ref{sLiouville}) can be recast into the customary Liouville equation \cite{Balescu},
\begin{equation}\label{Hdynamics}
\frac{\partial \rho}{\partial t} + \{\rho, \tilde H\} = 0
\end{equation}
where the effective Hamiltonian $\tilde H$ generates the conservative dynamics: \begin{eqnarray}
\dot q_i &=& \frac{\partial \tilde H}{\partial  p_i} = \{ q_i,\tilde H\}, \label{Ham1} \\
\dot p_i &=& - \frac{\partial \tilde H}{\partial q_i }= \{ p_i,\tilde H\} . \label{Ham2}
\end{eqnarray}

When the statistical correlation is taken into account, i.e. if $\varphi\neq 0$, the influence from the reservoir ${\cal R}_i$ cannot be represented as a potential force.
In this case, the reservoir force becomes momentum-dependent, in general, which we set
\begin{equation}\label{r-force1}
{\cal R}_i \equiv {\cal D}_i(q,p;T)
\end{equation}
in which we have indicated explicitly the dependence of ${\cal D}$ on the reservoir temperature $T$ in the canonical ensemble $\rho_R$.
Subsequently, Eq.~(\ref{sLiouville}) can be further manipulated to give
\begin{equation}\label{gLiouville}
\frac{\partial \rho}{\partial t} + \sum\left( \dot q_i\frac{\partial \rho}{\partial q_i} +
\dot p_i\frac{\partial \rho}{\partial p_i} \right) = - \Lambda\rho.
\end{equation}
The preceding equation describes a \textit{generalized} Liouville dynamics in the phase space of the system variables alone.
Note here that the second term on the left-hand-side (LHS) of Eq.~(\ref{gLiouville}) cannot be represented
as the Poisson bracket, $\{\rho, H\}$ because the dynamic variables evolve now according to
\begin{eqnarray*}
\dot q_i &=& \{q_i, H\}, \\
\dot p_i &=& \{p_i, H\} + {\cal D}_i
\end{eqnarray*}
which contain the effective, momentum-dependent force ${\cal D}$ arising from interaction between the system and the reservoir.
The factor $\Lambda$ on the right-hand-side (RHS) in Eq.~(\ref{gLiouville}), which is identified to be
\[
\Lambda(q,p;T) = \sum\frac{\partial}{\partial p_i}{{\cal D}_i}, \]
provides a measure of contraction, or expansion, of phase volume.

Now that we have treated the coordinate-dependent and momentum-dependent forces, we consider the external
time-dependent forces, frequently occurring when a mechanical or electromagnetic control of the system is required.
In the Jarzynski scheme such an external control is formally furnished with a time-dependent parameter, say $\lambda(t)$, in the system Hamiltonian as
\begin{equation}\label{sysHam}
H(q,p;\lambda)\equiv {\cal H}={\cal H}(q,p;\lambda).
\end{equation}
In this picture the prescribed Hamiltonian ${\cal H}(q,p;\lambda)$ develops non-autonomously in time because $\lambda$ is dynamically independent of $q$ and $p$, namely
\[ \frac{\partial \lambda(t)}{\partial q_i} = 0 = \frac{\partial \lambda(t)}{\partial p_i}.\]
The dynamical variables in ${\cal H}$ obey the \textit{extended} Hamilton equations of motion in the form,
\begin{eqnarray}
\dot q_i &=& \{q_i, {\cal H}\}, \label{gHam1} \\
\dot p_i &=& \{p_i, {\cal H}\} + {\cal D}_i(q,p;T), \label{gHam2}
\end{eqnarray}
which do not carry the manipulating forces.
Here, we emphasize the temperature dependence in ${\cal D}_i$ which features an essential difference between the deterministic Hamiltonian dynamics for a thermally open system and stochastic dynamics for a Brownian motion.
In the latter the frictional force is phenomenologically treated as independent of temperature.

Another picture attainable is to prescribe the external control directly in the equations of motion and to define the mechanical energy of the system as the instantaneous value of the bare Hamiltonian \cite{Taylor},
\begin{equation}\label{Ham}
H(q,p) = \sum\left\{ \frac{p^2}{2m} + V(q)\right\}
\end{equation}
where the potential energy $V(q)$ includes both interaction among constituents of the system and any other conservative external potentials.
We term the latter picture the \textit{system-centric} description in the sense that the energy of the system is specified solely by the system variables.
In order to gain some insight into how time-dependent forces enter the equations of motion, let us assume that the external control may be isolated as a perturbation $H^\prime$ to the bare Hamiltonian $H$.
Then, the system Hamiltonian is written additively as
\begin{equation}\label{sysHam2}
{\cal H}(q,p;\lambda(t)) =  H(q,p) + H^\prime(q,p;\lambda(t))
\end{equation}
where the perturbation term is not necessarily small.
With the preceding recipe for ${\cal H}$, Eqs.~(\ref{gHam1}) and (\ref{gHam2}) generate the extra terms ${\cal V}^{ex}$ and ${\cal G}^{ex}$ in the equations of motion,
\begin{equation}\label{extvel}
{\cal V}^{ex}_i \equiv \frac{\partial H^\prime}{\partial p_i}
\end{equation}
which contributes to time-development of the generalized velocity and
\begin{equation}\label{extfor}
{\cal G}^{ex}_i \equiv -\frac{\partial H^\prime}{\partial q_i}
\end{equation}
which describes the external, control force, acting on the degree of freedom $i$ in the open system.

Finally, we propose the extended Hamilton equations of motion in the system-centric picture as
\begin{eqnarray}
\dot q_i &=& \{ q_i, H\} + {\cal V}^{ex}_i(q,p;{\bf r},t), \label{extHam1} \\
\dot p_i &=& \{ p_i, H\} + {\cal D}_i(q,p;T) + {\cal G}^{ex}_i(q,p;{\bf r},t). \label{extHam2}
\end{eqnarray}
The dissipative force ${\cal D}$, stemming from the statistical correlation between the system and the reservoir, plays the role of a thermostat.
Although we have identified the external fields, ${\cal V}^{ex}$ and ${\cal G}^{ex}$ via Eqs.~(\ref{extfor}) and (\ref{extvel}), they are not, in general, derivable from a Hamiltonian.
They are required to take care of the coupling of the system to the external, nonpotential \textit{fields} at field point $\bf r$ in an open system in the system-centric view.
Equation~(\ref{extHam1}) suggests that the canonical momentum $p_i$ be not related to the generalized velocity ${\dot q}_i$ in the usual sense, but, under the extended dynamics, is given by
\[ p_i = m (\dot q_i - {\cal V}^{ex}_i).\]

Evidently, the proposed, extended equations of motion, Eqs.~(\ref{gHam1}) and (\ref{gHam2}) in the nonautonomous picture and Eqs.~(\ref{extHam1}) and (\ref{extHam2}) in the system-centric picture, constitute a non-Hamiltonian dynamics due to the non-potential terms.
Note also that the two Hamiltonians ${\cal H}$ and $H$ must be identical when the perturbation is turned off.
Some examples of such extended dynamics are considered in Sec.~\ref{Examples}.

\section{Mechanical work-energy theorems}
\label{mechanical work-energy}

The deterministic state of an open system may be depicted as a trajectory in phase space,
governed either by the nonautonomous Eqs.~(\ref{gHam1}) and (\ref{gHam2}) or by the system-centric
Eqs.~(\ref{extHam1}) and (\ref{extHam2}).
The mechanical energy of the system is specified as an instantaneous value of Eq.~(\ref{sysHam}) in the former or that of Eq.~(\ref{Ham}) in the latter, which is not conservative in either case.

We evaluate here how the energy of the system changes over a temporal interval $\tau$, evolving under the extended dynamics.
It can be done in the system-centric picture by carrying out the following manipulation of $H(q,p)$
given in Eq.~(\ref{Ham}),
\[
\Delta H = \int_0^\tau dt \dot H = \int_0^\tau dt\sum\left( \frac{\partial H}{\partial p_i} \dot p_i +
\frac{\partial H}{\partial q_i} \dot q_i \right).
\]
The required step is to substitute Eqs.~(\ref{extHam1}) and (\ref{extHam2}) for the time-rate
of the dynamical variables in the above expression.
Subsequently, it can be seen that the conservative dynamics is canceled out.
Then, the following identification from Eq.~(\ref{Ham}) is used in the remained terms,
\[
\frac{\partial H}{\partial p_i} = \frac{p_i}{m} \quad{\rm and}\quad \frac{\partial H}
{\partial q_i} = \frac{\partial V}{\partial q_i}. \]
Consequently, the induced change in the system-centric energy is represented as
\begin{equation}\label{work-energy}
\Delta H = \int_0^\tau dt \sum \frac{p_i}{m}{\cal D}_i + W
\end{equation}
where the first term on the RHS describes the energy dissipation into the surroundings by the momentum-dependent force ${\cal D}$.
The expression $W$ on the RHS of Eq.~(\ref{work-energy}) represents the work done by the external fields on the system,
\begin{equation}\label{control}
W \equiv \int_0^\tau dt\sum \left\{ \frac{p_i}{m} {\cal G}^{ex}_i + \frac{\partial V}{\partial q_i}{\cal V}^{ex}_i\right\}.
\end{equation}
When the external velocity-field ${\cal V}^{ex}$ is not coupled to the system, it holds from Eq.~(\ref{extHam1}) that the canonical momentum is related to the generalized velocity as usual, i.e. $p_i = m \dot q_i.$
The external work, then, is specified by the time-dependent force ${\cal G}^{ex}$ alone in its conventional form,
\begin{equation}\label{control2}
W = \sum \int_0^\tau dt \dot q_i {\cal G}^{ex}_i.
\end{equation}
Equation~(\ref{work-energy}) constitutes the conventional, \textit{mechanical} work-energy theorem in
the integral representation \cite{Taylor}, extended to accommodate the various sort of non-potential forces.
It explains transformation of the mechanical energy:
The mechanical energy $H$ increases with the external work $W$ performed on the system and decreases by the energy-exchange interaction ${\cal D}$ with the surroundings.

The work-energy theorem may be envisaged with the nonautonomous Hamiltonian ${\cal H}(p,q;\lambda)$, as well.
The dynamical variables in ${\cal H}$ obey the generalized equations of motion given in Eqs.~(\ref{gHam1})
and (\ref{gHam2}), while the parameter $\lambda(t)$ is manipulated according to a prescribed protocol
over the period $0\le t\le \tau$.
The induced change in ${\cal H}$ is readily evaluated as
\begin{eqnarray}\label{nonautoW-E}
\Delta {\cal H} &=& \int_0^\tau dt \dot {\cal H}\nonumber\\
&=& \int_0^\tau dt \sum \dot q_i{\cal D}_i + {\cal W}
\end{eqnarray}
where ${\cal W}$ is the parametric change of ${\cal H}(p,q;\lambda)$,
\begin{equation}\label{control3}
{\cal W} = \int_0^\tau dt \dot\lambda\frac{\partial {\cal H}}{\partial \lambda}.
\end{equation}
In passing to the second line in Eq.~(\ref{nonautoW-E}), the Hamiltonian dynamics has been canceled out
but the contribution from ${\cal D}_i$, the first term on the RHS.
The second term ${\cal W}$ on the RHS of Eq.~(\ref{nonautoW-E}) represents the control work done on the system by an external agent.
Equation~(\ref{nonautoW-E}) is the desired work-energy theorem pictured with the nonautonomous Hamiltonian ${\cal H}$.

Note that both work-energy theorems contain the same dissipation term; however, the definition of work
appears distinctively.
The two descriptions do not provide an equivalent measure to the mechanical energy of the system.
To clarify how the change induced in the nonautonomous Hamiltonian $\Delta {\cal H}$ differs from that in the system-centric Hamiltonian $\Delta H$, we use the additive perturbation model for the external manipulation, Eq.~(\ref{sysHam2}).
The change in the perturbation term $H^\prime$ over the work period can be calculated via
\[
\Delta H^\prime = \int_0^\tau dt \left\{\sum  \left( \frac{\partial H^\prime}{\partial q_i}\dot q_i + \frac{\partial H^\prime}{\partial p_i}\dot p_i \right) + \frac{\partial H^\prime}{\partial t}\right\}
\]
where $\partial H^\prime /\partial p_i$ and $\partial H^\prime /\partial q_i$ specify the external fields,
Eqs.~(\ref{extvel}) and (\ref{extfor}).
Then, $\Delta{\cal H}$ is obtained by adding the calculated $\Delta H^\prime$ to the energy change $\Delta H$ specified in Eq.~(\ref{work-energy}) as
\[
\Delta {\cal H} = \Delta H + \Delta H^\prime.
\]
The outcome has been shown to be exactly the same as the one given in Eq.~(\ref{nonautoW-E}).
In order to be more concrete, let us consider the linear coupling model,
\begin{equation}\label{dipole}
H^\prime = - \lambda(t) G(\{q_i\})
\end{equation}
where $G(\{q_i\})$ is a phase-observable which is conjugate to the control parameter $\lambda$.
The perturbation Hamiltonian describes a mechanically forced interaction or a dipole excitation in electromagnetic systems.
The external fields associated with the perturbation are identified immediately by
Eqs.~(\ref{extvel}) and (\ref{extfor}) as
\[
{\cal G}^{ex}_i = \lambda\frac{\partial G}{\partial q_i} \quad{\rm and}\quad {\cal V}^{ex}_i =0.
\]
Then, the work done on the system-centric Hamiltonian $H$ is written via Eq.~(\ref{control}) as
\begin{equation}\label{defwork1}
W = \sum \int {\cal G}^{ex}_i dq_i.
\end{equation}
On the other hand, it is given in the nonautonomous picture via Eq.~(\ref{control3}) as
\begin{equation}\label{defwork2}
{\cal W} = - \sum \int_0^\tau dt \dot{\cal G}^{ex}_i q_i = - \sum \int q_i d{\cal G}^{ex}_i.
\end{equation}
Thus, the two distinctive representations of external work have come to realization from the identical time-dependent force ${\cal G}^{ex}$:
In the system-centric description the work $W$, Eq.~(\ref{defwork1}) is represented as the integral of the forces
over the displacements of coordinates.
In contrast, the work ${\cal W}$, Eq.~(\ref{defwork2}) is represented in the nonautonomous description as
the negative integral of the coordinates over the variation of the forces.
The former is referred to as the exclusive work and the latter as the inclusive work by Jarzynski \cite{JarPhysique2007}.
By comparing two work-energy theorems, Eqs.~(\ref{work-energy}) and (\ref{nonautoW-E}), it follows that
\begin{equation}\label{work-diff}
{\cal W} = W + \Delta H^\prime
\end{equation}
where
$ \Delta H^\prime  = - \Delta \left[\sum q_i {\cal G}^{ex}_i\right].$ The preceding Eq.~(\ref{work-diff}) shows that the nonautonomous work ${\cal W}$ differs from the system-centric work $W$ by the amount of the induced energy from the perturbation.
Evidently, the energies defined by the two descriptions do not measure the same amount of quantity in an identical setup.

The nonautonomous description appears to carry a potential ambiguity in defining the physical energy-difference.
The reason is that the time-dependent Hamiltonian specifies the energy only up to an arbitrary time-dependent
factor without affecting the equations of motion \cite{Rubi2008}.
Consequently, the reference point of the energy may not be the same in initial and final states.
We argue, however, that the explicit time-dependence of ${\cal H}$ through a coupling mechanism of a macroscopic apparatus to the system [e.g. Eq.~(\ref{dipole})], is not to be introduced arbitrarily but in a macroscopically controllable manner.

\section{Statistical work-energy theorems}
\label{JarzynskiEquality}

Having established the mechanical work-energy theorems, we now proceed to formulate
their \textit{statistical} counterparts.
In a small system with a few degrees of freedom the individual, trajectory-dependent work may be an observable,
however, fluctuation involved in the work measurement hinders the mechanical work-energy theorem from being useful.
In performing work on a finite system instructed by a definite protocol, a myriad of trajectories participate due to insufficient information on the initial phase.
In both cases, a statistical description is required.

Here, we consider a finite system which is prepared initially in equilibrium with a surrounding reservoir
at temperature $T$ and subsequently undergoes NEQ transformation manipulated by an external control.
In the framework of classical statistical mechanics, later stage of the system is specified by a time-dependent ensemble density in phase space \cite{Balescu}.
We conceive that the system remains in contact with the heat reservoir, regardless of the coupling strength \cite{Jarzynski2004}, which seems natural in an experimental setup.

\subsection{Quasi-static average}

The external, control work induces a change in the energy of the system limited by the mechanical work-energy theorem.
When the work is performed quasi-statically, it may be assumed that the system remains in equilibrium with the heat reservoir.
Accordingly, the ensemble density $\rho(t)$ retains its canonical, equilibrium form $\rho_{eq}$ at each instant over the work interval, $0\le t \le \tau$.

Here, we attempt directly to take \textit{statistical average} of the work-energy balance over the quasi-equilibrium ensemble.
First, we perform the average in the system-centric picture, Eq.~({\ref{work-energy}), to evaluate
\begin{equation}\label{qsav}
\langle \Delta H \rangle_{eq} = \langle W\rangle_{eq} + \langle \int_0^\tau dt \sum \dot q_i{\cal D}_i \rangle_{eq}
\end{equation}
where $\rho_{eq}$ is specified by the system-centric Hamiltonian, Eq.~(\ref{Ham}) at each instant.
Under the quasi-static assumption, the average of $\Delta H$ may be evaluated by switching the order of the ensemble average and the time-integral as
\[
\langle \Delta H \rangle_{eq} = \langle \int_0^\tau dt \dot H \rangle_{eq} = \int_0^\tau d\langle H\rangle_{eq}.
\]
Then, it follows immediately that
\begin{equation}\label{qsav-energy}
\langle \Delta H \rangle_{eq} = U(\tau)-U(0) = 0
\end{equation}
where $U$ is the \textit{internal} energy defined to be
\[
U = \int dqdp~ \rho_{eq}H(q,p).
\]
The second equality in Eq.~(\ref{qsav-energy}) arises because the functional form of $H(q,p)$ remains the same at each instant and the dynamic variables span the entire phase space.
Consequently, the internal energy remains to be constant in quasi-static processes.
Similarly, the average of the external work $W$ in Eq.~(\ref{qsav}) can be performed by resorting to the explicit representation, Eq.~(\ref{control2}), neglecting ${\cal V}^{ex}$ here, as
\begin{eqnarray*}
\langle W\rangle_{eq} &=& \int dqdp \rho_{eq}(t) \sum \int_0^\tau dt {\dot q}_i {\cal G}^{ex}_i \\
&=& \int_0^\tau dt \langle\sum {\dot q}_i {{\cal G}}^{ex}_i\rangle_{eq}.
\end{eqnarray*}
The integrand in the preceding expression is the averaged power,
\[
\frac{d W_{th}}{dt} = \langle\sum {\dot q}_i {{\cal G}}^{ex}_i\rangle_{eq}.
\]
Accordingly, one can write
\begin{equation}\label{qsav-work}
\langle W\rangle_{eq} = \int dW_{th} \equiv W_{th}
\end{equation}
where $dW_{th}$ is the infinitesimal \textit{thermodynamic work} done on the system
which is not an exact differential.
Lastly, the statistical average of the dissipation term on the RHS of Eq.~(\ref{qsav}) may be evaluated as
\begin{eqnarray*}
&& \langle \int_0^\tau dt \sum \dot q_i{\cal D}_i \rangle_{eq} \\
&=& \int_0^\tau dt  \int dqdp \rho_{eq}(t) \sum \dot q_i{\cal D}_i \\
&=& - \int_0^\tau dt \langle \Theta(q,p;\beta)\rangle_{eq}
\end{eqnarray*}
where $\Theta$ is the Rayleigh dissipation function defined to be \cite{Pars}
\begin{equation}\label{Rayleigh}
\Theta\equiv - \sum{\cal D}_i \dot q_i.
\end{equation}
Here, we identify the heat $Q$ absorbed into the system, when its sign is positive, as
\begin{equation}\label{qsav-heat}
Q  \equiv  -\int_0^\tau dt \langle \Theta(q,p;\beta,t)\rangle_{eq} = \int TdS
\end{equation}
where $S$ represents the Clausius entropy of the system.
Finally, by collecting the obtained expressions, Eqs.~(\ref{qsav-energy}), (\ref{qsav-work}), and (\ref{qsav-heat}) into Eq.~(\ref{qsav}), we reach the first law of thermodynamics in the system-centric picture,
\begin{equation}\label{firstlaw}
W_{th} + Q =0
\end{equation}
which states that the work done on the system is precisely balanced with the heat dissipated into the surroundings.
The internal energy is unchanged in the system-centric description.

Next, we carry on statistical average in the nonautonomous picture, Eq.~(\ref{nonautoW-E}), which takes
\begin{equation}\label{qsav2}
\langle \Delta {\cal H} \rangle_{eq} = \langle {\cal W}\rangle_{eq} + \langle \int_0^\tau dt \sum \dot q_i{\cal D}_i\rangle_{eq}
\end{equation}
where $\rho_{eq}$ is specified in terms of the parametric Hamiltonian, Eq.~(\ref{sysHam}).
Unlike Eq.~(\ref{qsav-energy}), the change in the internal energy is not zero in the time-dependent energy landscape,
\begin{equation}\label{qsav-energy2}
\langle \Delta {\cal H} \rangle_{eq} = U(\lambda(\tau))-U(\lambda(0)) = \Delta U,
\end{equation}
because $U$ differs at each quasi-static instant,
\[
U(\lambda(t)) = \int dqdp~ \rho_{eq}{\cal H}(p,q;\lambda(t)).
\]
The average of the parametric work Eq.~(\ref{control3}) can be manipulated as
\begin{eqnarray*}
\langle {\cal W}\rangle_{eq} &=& \langle \int_0^\tau dt \dot\lambda\frac{\partial {\cal H}}{\partial \lambda}\rangle_{eq} \\
&=& \langle \int d\lambda \frac{\partial {\cal H}}{\partial \lambda} \rangle_{eq} \\
&=& \int d\lambda \frac{\partial}{\partial \lambda} \langle {\cal H}\rangle_{eq}.
\end{eqnarray*}
In passing to the third line in the preceding manipulation, use has been made of the fact that the parameter is not a coordinate in phase space.
Then, the thermodynamic work ${\cal W}_{th}$ is represented as
\begin{equation}\label{qsav-work2}
{\cal W}_{th} = \int d\lambda \frac{\partial U(\lambda)}{\partial \lambda}.
\end{equation}
The physical representation of the heat remains the same with the system-centric picture, Eq.~(\ref{qsav-heat}).
Consequently, the first law takes the conventional form as
\begin{equation}\label{firstlaw2}
\Delta U = {\cal W}_{th} + Q
\end{equation}
which includes the change in the internal energy for an open system.

Equation~(\ref{firstlaw2}) can be recast in terms of the equilibrium Helmholtz FE, defined as $F=U-TS$, after a Legendre transformation,
into the form \cite{Huang}
\[
\Delta F = {\cal W}_{th} - \int SdT
\]
which for an isothermal transformation reduces to
\begin{equation}\label{isoth1stlaw}
\Delta F = {\cal W}_{th}.
\end{equation} Note that all thermodynamic variables maintain their quasi-equilibrium states over the work period and that the induced transformation is \textit{reversible}, in the quasi-static processes.
We also emphasize that the first law must hold not only in reversible processes but also in irreversible processes, albeit we have derived it only in the quasi-static limit.

\subsection{NEQ ensemble average}
\label{secNEQenav}

Under general NEQ conditions, the ensemble density of the system does not keep up its quasi-equilibrium state continually over the work period.
Instead, the NEQ ensemble density $\rho(t)$ is governed by the \textit{generalized} Liouville equation in phase space, Eq.~(\ref{gLiouville}) which is rewritten here for convenience as
\begin{equation}\label{GLiouville}
D_t\rho(t)= - \Lambda\rho
\end{equation}
where $D_t$ is the convective derivative along the phase flow,
\[ D_t\equiv \frac{\partial}{\partial t} + \sum\left( \dot q_i\frac{\partial}{\partial q_i} +
\dot p_i\frac{\partial}{\partial p_i} \right).\]
The function $\Lambda$ on the RHS of Eq.~(\ref{GLiouville}) is the compressibility factor defined as a divergence in phase space,
\begin{equation}\label{compfactor}
\Lambda(q,p;t) = \sum\left( \frac{\partial}{\partial {q_i}}\dot{q_i} + \frac{\partial}{\partial p_i}\dot{p_i} \right).
\end{equation}
The dynamical variables $q_i$ and $p_i$ obey either Eqs.~(\ref{gHam1}) and (\ref{gHam2}) or Eqs.~(\ref{extHam1}) and (\ref{extHam2}), depending on the choice of the energy picture.
Note that the second term in the convective derivative $D_t\rho$ cannot be written as a Poisson bracket in the extended Hamilton dynamics.

Although it is not tractable to solve Eq.~(\ref{GLiouville}), a formal solution can be written by a direct integration in the moving frame with the phase fluid.
The result is given as
\begin{equation}\label{tdensity}
\rho(q,p;t) = {\cal U}(t) \rho(q,p;0)
\end{equation}
where ${\cal U}$ is the time-evolution operator defined by
\begin{equation}\label{propagator}
{\cal U}(t) = \exp\left\{-\int_0^t dt^\prime \Lambda(q,p;t^\prime)\right\}
\end{equation}
and $\rho(q,p;0)$ represents an initial density at $t=0$.
Equation~(\ref{tdensity}) manifests that the compressibility function $\Lambda$ plays the role of a generator of \textit{convective time-translation} in the present formulation.
In the ordinary Liouvillean dynamics which preserves the phase volume, the NEQ density becomes a constant of motion, i.e. $D_t\rho=0$.

The initial equilibrium states are characterized by the same density both in the system-centric and nonautonomous descriptions because the nonautonomous Hamiltonian before turning on the work parameter is identical to the system-centric Hamiltonian,
\[{\cal H}(q,p;0)=H(q_0,p_0).\]
In both energy pictures the initial density is physically specified as the canonical equilibrium distribution,
\begin{eqnarray}\label{canonical}
\rho(q,p;0 ) &=& \rho_{eq}(q_0,p_0) \nonumber\\
&=& \exp\{\beta (F_0-H(p_0,q_0))\}
\end{eqnarray}
where $\beta=1/(k_BT)$, $k_B$ and $T$ being the Boltzmann constant and the temperature of the heat reservoir, respectively, and $F_0$ is the Helmholtz FE at initial stage.

We now proceed to evaluate the statistical average of the mechanical work-energy theorem when
the external work causes a fast change in the system during the work period $\tau$.
We shall consider the problem in the system-centric description first.
Instead of directly taking the average of Eq.~(\ref{work-energy}), however, we adopt the Jarzynski construction
which defines the average of control work as a NEQ ensemble average of the \textit{weighted work-exponential} in phase space.
Technically, the average at $t=\tau$ may be taken equivalently either in the Schr{\"o}dinger picture or in the Heisenberg picture in quantum mechanical terms, which is elaborated below. When the former is employed, the average is taken over the time-dependent distribution
while the exponentiated work is considered a fixed-time phase function as
\begin{equation}\label{Schrodinger}
\langle e^{-\beta W} \rangle  = \int dqdp e^{-\beta W}\rho(q,p;\tau).
\end{equation}

It is important here to recognize that the phase-space measure $dqdp$ is not invariant in the extended Hamiltonian dynamics.
Rather, the two generalized Liouville measures $dq_0dp_0$ and $dqdp$ at different times $t=0$ and $\tau$, respectively, are related to each other via the Jacobian ${\cal J}$,
\begin{equation}\label{Lmeasure}
dq dp = {\cal J}(\tau)dq_0dp_0.
\end{equation}
The preceding Jacobian ${\cal J}$ itself obeys dynamics in the opposite sense to Eq.~(\ref{GLiouville}})
\[ D_t{\cal J} = \Lambda{\cal J}.\]
For the given initial condition, ${\cal J}(0)=1$, it can be formally integrated to give \cite{Arnold}
\begin{equation}\label{Jacobian}
{\cal J}(\tau) = \exp\left\{\int_0^\tau dt \Lambda(q,p;t)\right\}.
\end{equation}
By inspecting that the Jacobian ${\cal J}$ differs from the time-evolution operator ${\cal U}$ only by the sign of the exponent, we attain that ${\cal J}$ and ${\cal U}$ evolve in the inverse sense to each other,
\begin{equation} \label{unitarity}
{\cal J}(\tau) {\cal U}(\tau) =1.
\end{equation}
The preceding equation is the classical-mechanical \textit{unitarity} condition which imposes the preservation of ensemble members in phase space,
\begin{eqnarray*}
&&\int dqdp \rho(q,p;t) \\
=&& \int \left\{{\cal J} dq_0 dp_0\right\}\left\{ {\cal U}\rho(q,p;0)\right\} \\
=&& \int dq_0 dp_0 \rho(q,p;0).
\end{eqnarray*}

With help of Eqs.~(\ref{tdensity}) and (\ref{Lmeasure}), Eq.~(\ref{Schrodinger}) can be rewritten as
\begin{equation}\label{Heisenberg}
\langle e^{-\beta W} \rangle = \int dq_0dp_0 e^{-\beta W(\tau)} \rho_{eq}(q,p;0)
\end{equation}
where the subsequent rearrangement has been made of
\[ e^{\int_0^\tau dt \Lambda(q,p;t)} e^{-\beta W} e^{-\int_0^\tau dt \Lambda(q,p;t)} = e^{-\beta W(\tau)}.\]
Equation~(\ref{Heisenberg}) is the Heisenberg representation of the NEQ ensemble average where the average is taken over the initial equilibrium ensemble, whereas the work function is interpreted to have evolved over the time-interval $\tau$, limited by the mechanical work-energy theorem, Eq.~(\ref{work-energy}).
The NEQ averages may be viewed as a functional which maps the phase function of the work exponential
onto a scalar in phase space.
We just proved that the two pictures are identical in carrying out the NE averages.

Here, we carry on our calculation in the Schr\"odinger picture to substitute the NEQ ensemble density, Eq.~(\ref{tdensity}) at $t=\tau$ for $\rho(p,q;\tau)$ in Eq.~(\ref{Schrodinger}).
Conforming to the mechanical work-energy theorem Eq.~(\ref{work-energy}), we replace the control work $W$ with the energy gain $\Delta H$ minus the dissipated energy in Eq.~(\ref{Schrodinger}) to cast the work exponential into
\begin{widetext}
\begin{equation}\label{workexp}
e^{-\beta W} =  \exp\left[-\beta\left\{ H(q,p;\tau) - H(q_0,p_0) - \int_0^\tau dt \sum \frac{p_i}{m} {\cal D}_i
\right\}\right].
\end{equation}
\end{widetext}
Then, it is straightforward to rearrange the integrand on the RHS of Eq.~(\ref{Schrodinger}) to bring about
\begin{widetext}
\begin{equation}\label{precursor}
\langle e^{-\beta W} \rangle = e^{\beta {F_0}}\int dq dp \left\{{\cal J}^{-1}
e^{\beta\int_0^\tau dt \sum{{\cal D}_i}p_i/m}\right\}
e^{-\beta H(q,p;\tau)}
\end{equation}
\end{widetext}
where ${\cal J}^{-1}$ is the inverse Jacobian.
Now, one can observe that if the expression in curly brackets in the integrand on the RHS of Eq.~(\ref{precursor}) reduces to unity, i.e.
\begin{equation}\label{criterion}
{\cal J}^{-1}e^{\beta\int_0^\tau dt \sum{{\cal D}_i}p_i/m} = 1,
\end{equation}
then Eq.~(\ref{precursor}) turns into
\[
\langle e^{-\beta W} \rangle =  e^{\beta {F_0}} e^{-\beta {F(\tau)}}
\]
where $F(\tau)$ is the Helmholtz FE at $t=\tau$,
\[ F(\tau) = -\beta^{-1} \ln\left\{ \int dq dp e^{-\beta H(q,p;\tau)}\right\}.\]
The value of the system-centric Hamiltonian varies with time as a function of the dynamical variables, but the functional  form of $H(q,p)$ is fixed at each instant.
Accordingly, the resulting FE is the same at the initial and final equilibrium states with the identical temperature $\beta$, $F(\tau)=F(0)$.
Consequently, the average of the exponential work becomes
\begin{equation}\label{Bochkov}
\langle e^{-\beta W} \rangle = 1
\end{equation}
which is the Bochkov-Kuzovlev work relation \cite{Bochkov1977,JarPhysique2007}.

We now turn our attention to formulating the NEQ work theorem in the nonautonomous picture.
To this end, we only need to use the alternative representation of the mechanical work-energy theorem, Eq.~(\ref{nonautoW-E}), in defining the work exponential Eq.~(\ref{workexp}).
As previously mentioned, the initial ensemble is identical to the system-centric case.
Then, the statistical average Eq.~(\ref{precursor}) is replaced by
\begin{widetext}
\begin{equation}\label{precursor2}
\langle e^{-\beta {\cal W}} \rangle = e^{\beta {F_0}}\int dq dp \left\{{\cal J}^{-1}
e^{\beta\int_0^\tau dt \sum{{\cal D}_i}p_i/m}\right\}
e^{-\beta {\cal H}(q,p;\lambda(\tau))}.
\end{equation}
\end{widetext}
It is evident that Eq.~(\ref{precursor2}) becomes the proclaimed Jarzynski equality \cite{JarPRL1997},
\begin{equation}\label{JE}
\langle e^{-\beta {\cal W}} \rangle = e^{-\beta\Delta {\cal F}}
\end{equation}
where $\Delta {\cal F} = {\cal F}(\tau)-F_0$
if the same condition given in Eq.~(\ref{criterion}) meets.
In this case, however, ${\cal F}(\tau) \neq F_0$ because the instantaneous FE (${\cal F}$) depends not only on the reservoir temperature but also on the control parameter $\lambda$,
\[ {\cal F}(\beta,\lambda) = -\beta^{-1} \ln\left\{ \int dq dp e^{-\beta {\cal H}(q,p;\lambda)}\right\}.\]

The derived restraint Eq.~(\ref{criterion}) is physically satisfied when
\begin{equation}\label{criterion2}
\int_0^\tau dt\Lambda(q,p;\beta,t) + \beta \int_0^\tau dt \Theta(q,p;\beta,t) \equiv 0
\end{equation}
where $\Theta$ is the Rayleigh dissipation function previously defined in Eq.~(\ref{Rayleigh}).
Note that we have indicated the temperature dependence explicitly both in the compressibility factor $\Lambda$ and in the Rayleigh function $\Theta$, via ${\cal D}$ Eq.~(\ref{r-force1}) in their definitions.
The condition Eq.~(\ref{criterion2}) has been derived without invoking any specific models.
It asserts that the JE is loosened unless the non-vanishing $\Lambda$ and the scaled, dissipative power $\beta\Theta$ cancel exactly each other out over the NEQ work performance in a thermally open system.

The compressibility factor $\Lambda$, defined in Eq.~(\ref{compfactor}), consists of three parts in the system-centric picture,
\begin{eqnarray}\label{gencompr}
\Lambda &=& \sum\left\{ \frac{\partial}{\partial p_i} \left( {\cal D}_i + {\cal G}^{ex}_i  \right) +
\frac{\partial}{\partial q_i} {\cal V}^{ex}_i \right\} \nonumber\\
&\equiv & \Lambda_{\cal D} + \Lambda_{\cal G} + \Lambda_{\cal V},
\end{eqnarray}
where $\Lambda_{\cal D}$ is the contribution from the dissipation ${\cal D}$, $\Lambda_{\cal G}$
from the control force ${\cal G}^{ex}$, and $\Lambda_{\cal V}$ from the macroscopic velocity-coupling ${\cal V}^{ex}$.
In the nonautonomous Hamiltonian description, the compressibility factor takes only a single term,
\begin{equation}\label{gencompr2}
\Lambda = \Lambda_{\cal D}.
\end{equation}

Here, we discuss the physical implication of the enunciated condition given in Eq.~(\ref{criterion2}).
The finite system that we consider is assumed to remain in thermal contact with a single surrounding reservoir.
Therefore, before turning on or after turning off the work parameter $\lambda$ should the system come to equilibrium with the reservoir due to boundary interaction of the system with the thermal reservoir.
In this situation the generalized Liouville equation Eq.~(\ref{GLiouville}) must admit the canonical ensemble density $\rho_{eq}$ as its solution, which yields
\begin{equation}\label{criterion3}
\sum\left( \dot q_i\frac{\partial  \rho_{eq}}{\partial q_i} +
\dot p_i\frac{\partial  \rho_{eq}}{\partial p_i} \right) = -\Lambda \rho_{eq}
\end{equation}
which constitutes the \textit{detailed balance} in thermostatted dynamics.
By directly substituting Eq.~(\ref{canonical}) for $\rho_{eq}$ and, then, by making use of the extended Hamilton equations of motion excluding the external forces,
one can show that the above Eq.~(\ref{criterion3}) is reduced to
\[
\Lambda = -\beta\Theta
\]
which evidently satisfies Eq.~(\ref{criterion2}).
Both $\Lambda$ and $\Theta$ in Eq.~(\ref{criterion2}) are associated with the momentum-dependent force ${\cal D}$, Eq.~(\ref{r-force1}).
Therefore, Eq.~(\ref{criterion2}) implies essentially a consistency condition that the effective force ${\cal D}$, originating from the coarse specification of the boundary interaction by a macroscopic parameter $\beta$, must meet in order to assure the detailed balance condition.
The temperature-dependence of ${\cal D}$ provides the NEQ dynamics of a thermally open system with a thermostatting mechanism.

The JE, Eq.~(\ref{JE}) is the desired, statistical work-energy theorem applying to general NEQ processes beyond the quasi-static limit.
Appealing to Jensen's inequality \cite{Jensen},
\[ \langle e^{-\beta {\cal W}} \rangle \ge e^{-\beta \langle {\cal W}\rangle},\]
the NEQ work-energy theorem implies that there exists an excess of work, \textit{on average}, compared to the FE increment, i.e.
\begin{equation}\label{isoth2ndlaw}
\Delta {\cal F} \le \langle {\cal W}\rangle
\end{equation}
where the equality holds for an isothermal, quasi-static process, Eq.~(\ref{isoth1stlaw}).
In a small system with a few degrees of freedom it is not surprising to anticipate a statistical deviation from Eq.~(\ref{isoth2ndlaw}) that such an \textit{individual} mechanical process as $\Delta {\cal F} \ge W$ may occur occasionally.
However, it would not occur in a finite system on any realistic time-scale because the observable is the \textit{averaged} work, ${\cal W}_{th}=\langle {\cal W}\rangle$, not the individual realization of $W$, which is subsumed in the second law of thermodynamics in its general form \cite{Huang}
\begin{equation}\label{secondlaw}
\Delta {\cal F} \le {\cal W}_{th} - \int SdT.
\end{equation}
It is suggestive to observe that Eq.~(\ref{secondlaw}) tends to Eq.~(\ref{isoth2ndlaw}) in an \textit{isothermal} limit, implying that the validity of the JE may be restricted approximately to isothermal processes.

Finally, we want to mention a suggestive report by others where it is shown that the \textit{excess} of thermodynamic work, ${\cal W}_{th} - \Delta {\cal F}$ over the work period $\tau$ is bounded from below by an information-theoretic measure \cite{Vaikuntanathan2009}.
The measure is quantified as the relative entropy between the actual NEQ density $\rho(\tau)$ and the quasi-static equilibrium density $\rho_{eq}(\tau)$.

\section{Symmetric work fluctuation: The mesoscopic reversibility}
\label{CrooksTheorem}

Here, we explore the physical ground of the symmetric nature of the work FTs and its relation to the essentially one-way theorem of the JE.
We shall first consider the problem in the system-centric picture and continually describe the companion result from the nonautonomous picture.

To this end, it is essential to deduce under what conditions the generalized Liouville dynamics, Eq.~(\ref{GLiouville}) governed by the extended Hamilton equations of motion, Eqs.~(\ref{extHam1}) and (\ref{extHam2}), may be invariant under time (motion) reversal.
The time-reversal operation, denoted by ${\cal T}: t\rightarrow -t$, is formalized by the following discrete transformation:
\[ q\rightarrow q\quad {\rm and}\quad p\rightarrow - p.\]
Conforming to them, we postulate that to every density $\rho(q,p;t)$ at instant $t$ there corresponds a time-reversed density defined by
\[ {\cal T}\rho(q,p;t){\cal T}^{-1} = \rho(q,-p;-t).\]
Then, by inspecting Eq.~(\ref{GLiouville}), one can verify that
$\rho(q,-p;-t)$ is also solution to the generalized Liouville equation if the
compressibility factor changes its sign under time-reversal, i.e.
\begin{equation}\label{trevcomp}
 {\cal T} \Lambda (q,p;t){\cal T}^{-1} = - \Lambda(q,p;t).
\end{equation}
The preceding Eq.~(\ref{trevcomp}) is the required condition which makes the generalized Liouville
dynamics \textit{invariant} under time-reversal.
When it is satisfied, the time-reversed density propagates backward in time under the influence of
the propagator ${\cal U}$ as
\begin{equation}\label{trevdensity}
\rho(q,-p;-t) = {\cal U}(t) \rho(q_0,-p_0;0).
\end{equation}
The concrete representation, Eq.~(\ref{gencompr}) of $\Lambda$ under the extended Hamiltonian dynamics leads to
the physical conditions to be imposed on the external fields,
\begin{eqnarray}
{\cal G}_i^{ex}(-t) &=& {\cal G}_i^{ex}(t), \label{symcond-exF}  \\
{\cal V}_i^{ex}(-t) &=& -{\cal V}_i^{ex}(t),\label{symcond-exV} \end{eqnarray}
and on the nonconservative thermostatting force,
\begin{equation}\label{mesorev1}
{\cal D}(-p)={\cal D}(p)
\end{equation}
which states that the momentum-dependent force must be even under inversion, $p\rightarrow -p$,

The invariance condition Eq.~(\ref{trevcomp}) is special because the dynamics in generic NEQ work-measurements would be asymmetric, i.e. irreversible, in general.
We shall call the exploited symmetry of the generalized Liouville dynamics a \textit{dynamically mesoscopic} reversibility in the sense that the external, non-potential couplings in the extended Hamilton equations of motion, Eqs.~(\ref{extHam1}) and (\ref{extHam2}), are not microscopic but rather statistical in origin.
In particular, we have recapitulated in Sec.~\ref{ExEqMotions} that insufficient knowledge about the interaction of the system with the surroundings is represented as the momentum-dependent forces ${\cal D}$ on the system.
In below, we establish that the NEQ work fluctuation theorems in fact reflects the mesoscopic reversibility of
specially prepared dynamics.

To proceed, let us denote the two equilibrium states of the system by $A$ and $B$, respectively, at both ends
connected by a pre-determined work protocol over the duration $\tau$.
Corresponding to the system prepared in canonical equilibrium $\rho_{eq}(q,p;A)$ at $t=0$, the number of initial micro-states in the range $(q_0,q_0+dq_0)$ and $(p_0,p_0+dp_0)$
would be proportional to $\rho_{eq}(q,p;A)dq_0dp_0$.
Among these phase-space points, the number of initial micro-states belonging to a specific realization of forward work $W_F=W$ is restricted to $\rho_{eq}(q,p;A)dq_0dp_0 \delta(W_F-W)$
where $\delta(W_F-W)$ is the Dirac delta function. Thus, the work distribution $g_F$ in the forward process of performing work by an amount of $W$ may be written as
\begin{equation}\label{Fwdist}
g_F(W_F=W) = \int dq_0dp_0 \rho_{eq}(q,p;A)\delta(W_F-W)
\end{equation}
with normalization,
\[ \int g_F(W_F) dW_F = 1.\]
When the preceding distribution is used to evaluate the Jarzynski work average, it leads to
\begin{eqnarray*}
\langle e^{-\beta W}\rangle &\equiv &\int dW_F e^{-\beta W_F}g(W_F)\\
&=& \int dq_0dp_0  e^{-\beta W} \rho_{eq}(q,p;A)
\end{eqnarray*}
which is Eq.~(\ref{Heisenberg}) in the Heisenberg picture.

Similarly, one can construct the work distribution in the reversed process, $g_R$.
Such a reversed procedure is not permitted in general unless the forward work has been performed quasi-statically.
For a fast work process, we assume that the invariance condition Eq.~(\ref{trevcomp}) is enforced so that motion is still symmetric under time-reversal.
In this case the control work is pretended to be carried out precisely along backward trajectories by the amount of
\[W_R = {\cal T}W_F{\cal T}^{-1} = -W.\]
The reverse work-protocol sets up a new starting density as the time-reversed, ending equilibrium
state, $\rho_{eq}(q,p;B)$, from the forward process and allows the system to evolve
with abiding by Eq.~(\ref{trevdensity}).
Then, by the equivalent arguments used in specifying Eq.~(\ref{Fwdist}) the reverse work-distribution may be formulated as
\begin{widetext}
\begin{equation}\label{Rvdist}
g_R(W_R=-W) = \int {\cal T}[dqdp]{\cal T}^{-1} {\cal T}[\rho_{eq}(q,p;B)]{\cal T}^{-1}\delta(W_R-(-W)).
\end{equation}
\end{widetext}
The generalized Liouville measure $dqdp$ is invariant under time-reversal, ${\cal T}[dqdp]{\cal T}^{-1} = dqdp$.
The time-reversed equilibrium density at $B$ is given by
\[
{\cal T}[\rho_{eq}(q,p;B)]{\cal T}^{-1} = e^{\beta F_B} e^{-\beta {\cal T}[H(q,p;B)]{\cal T}^{-1}}
\]
where the time-reversed Hamiltonian is limited by the mechanical work-energy theorem Eq.~(\ref{work-energy}) but
in a temporally backward manner,
\begin{widetext}
\[
{\cal T}[H(A)-H(B)]{\cal T}^{-1} = {\cal T}W_R{\cal T}^{-1} + {\cal T}\left[\int_B^A dt\sum\frac{p_i}{m}
{\cal D}_i\right]{\cal T}^{-1}.
\]
\end{widetext}
The system-centric Hamiltonians are invariant under time-reversal and the control work considered is reversible with change of its sign as $W_R=-W_F$.
In addition, the dissipative work changes its sign under time-reversal by the imposed symmetry,
Eq.~(\ref{mesorev1}) on the thermostatting force, which has been enforced by the invariance condition Eq.~(\ref{trevcomp}).
Consequently, it turns out that the work-energy theorem applied in the backward sense is transformed, under time reversal, into the work-energy theorem in the forward direction,
\begin{widetext}
\[
H(q,p;B) = H(q_0,p_0;A) + W_F + \int_A^B dt\sum\frac{p_i}{m}{\cal D}_i. \]
\end{widetext}
We just verified that the mechanical work-energy theorem, Eq.~(\ref{work-energy}), also acts symmetrically under
time inversion in a mesoscopically reversible system.
By substituting the last expression into Eq.~(\ref{Rvdist}), one can obtain that
\begin{widetext}
\begin{eqnarray}\label{Rvdist1}
g_R(W_R) &=& e^{\beta F_B} e^{-\beta W}\int dqdp e^{- \beta \left\{H(q_0,p_0;A) +
\int_0^\tau dt \sum{\cal D}_i p_i/m \right\}}\delta(W_F-W) \nonumber \\
&=& e^{\beta F_B} e^{-\beta W} \int dq_0dp_0 \left\{ {\cal J} e^{  -\beta \int_0^\tau dt
\sum{\cal D}_i p_i/m } \right\} e^{-\beta H(q_0,p_0;A)} \delta(W_F-W)
\end{eqnarray}
\end{widetext}
where in the second step we have used the Jacobian relation, Eq.~(\ref{Lmeasure}).
Here, one can notice that the enclosed expression in curly brackets in Eq.~(\ref{Rvdist1}) is exactly what appears
in the enunciated consistency criterion for the NEQ work theorems, Eq.~(\ref{criterion}).
For such a work measurement satisfying the criterion, Eq.~(\ref{Rvdist1}) reduces to
\begin{equation}\label{BKWFT}
g_R(-W) = e^{-\beta W}g_F(W)
\end{equation}
which is the Bochkov-Kuzovlev version of the work FT \cite{Bochkov1977,JarPhysique2007}.

Next, we summarize the outcome from the nonautonomous picture, that we would have obtained instead of Eq.~({\ref{BKWFT}) if we had formulated with the alternative mechanical work-energy theorem,
Eq.~(\ref{nonautoW-E}).                           After imposing the consistency criterion, Eq.~(\ref{criterion}), on the companion expression to Eq.~(\ref{Rvdist1}), it can be shown straightforwardly that the result takes the form,
\begin{equation}\label{Crooks}
g_R(-{\cal W}) = e^{\beta (\Delta {\cal F}-{\cal W})}g_F({\cal W})
\end{equation}
with $\Delta {\cal F} = {\cal F}_B-{\cal F}_A$, which is the desired CWFT \cite{Crooks1998,Crooks1999}.
The required time-reversal constraints in deriving Eq.~(\ref{Crooks}}) are the invariance of the nonautonomous Hamiltonian,
\begin{equation}\label{trev2}
{\cal T}{\cal H}(q,p;\lambda){\cal T}^{-1} = {\cal H}(q,p;\lambda)
\end{equation}
and the already prescribed symmetry condition, Eq.~(\ref{mesorev1}).
The two conditions guarantee that the NEQ work is reversible, ${\cal W}_R=-{\cal W}_F$.
The JE, Eq.~(\ref{JE}), follows from the CWFT when both sides of Eq.~(\ref{Crooks}) are integrated
over the full rage of work values ${\cal W}$, assuming $g_F({\cal W})$ and $g_R(-{\cal W})$ are properly normalized.
However, the former is more general in applicability than the latter because it has been derived in Sec.~\ref{secNEQenav} without requiring the time-reversal invariance of the underlying dynamics.
Note that in typical single small-system experiments of testing the JE, the system must be brought back to
initial state after completing the unidirectional work \cite{Liphardt2002}.
It means that the dynamical reversibility of the system is still implicitly imposed on the actual realization of the JE.
The genuine irreversibility seems awaiting further to be explored.
We observe researchers have put forth an effort lately to extend the work FTs to account for irreversible transitions between partial equilibrium states \cite{Maragakis2008,Junier2009}.

\section{Examples}
\label{Examples}

We consider here a few examples to demonstrate how the consistency condition [Eq.~(\ref{criterion2})] for the NEQ work-energy theorems may be employed in actual NEQ dynamics.

\subsection{Isolated systems under time-dependent external forces}
\label{Isolated}

As a simple situation, let us consider that only a time-dependent manipulation is put into action
by an external agent on an, otherwise, isolated system.

In the system-centric picture, the system is described by the extended Hamilton equations of motion,
\begin{eqnarray*}
\dot q_i &=& \frac{1}{m} p_i, \\
\dot p_i &=& -\frac{\partial V}{\partial q_i} + {\cal G}^{ex}_i,
\end{eqnarray*}
in the time-independent, energy landscape, Eq.~(\ref{Ham}).
One can immediately see that the validity condition, Eq.~(\ref{criterion}) is satisfied because the control force does not contribute to the phase-space compressibility,
\begin{equation}\label{LambdaC}
\Lambda_{\cal G} = \sum \frac{\partial}{\partial p_i}{\cal G}_i^{ex}(t) = 0 .
\end{equation}
The conclusion is unchanged even if there is an additional dependence of the external force on the generalized
coordinates, ${\cal G}_i^{ex}(t)={\cal G}_i^{ex}(q;t)$.
Consequently, the Bochkov-Kuzovlev equality, Eq.~(\ref{Bochkov}) holds trivially.
On the other hand, in order to realize the symmetric work FT, Eq.~(\ref{BKWFT}) the external force must
further comply with the time-reversal symmetry, Eq.~(\ref{symcond-exF}). When the mesoscopic reversibility is satisfied, the reverse work is the negative of the forward work as
\begin{eqnarray}
W_F& =& \sum\int_0^\tau dt \dot q_i(t) {\cal G}_i^{ex}(t)\nonumber\\
& =& - \sum\int_0^\tau dt \dot q_i(\tau-t){\cal G}_i^{ex}(\tau-t)\nonumber\\
& =& - W_R.
\end{eqnarray}
This was stated formally using the time-reversal operator previously in Sec,~\ref{CrooksTheorem}.

In the nonautonomous Hamiltonian picture, Eq.~(\ref{sysHam2}), the external manipulation of the system must be built into the Hamiltonian as a time-dependent parameter.
For instance, single-molecule pulling experiments are often described by the phenomenological harmonic term \cite{Chen2010,Ritort2014},
\begin{equation} \label{harmonic}
H^\prime = \frac{1}{2}k\left\{G(\left\{q_i\right\})-\lambda(t)\right\}^2
\end{equation}
where $k$ is the spring constant and $G(\left\{q_i\right\})$ is the molecular extension which is a function of the generalized coordinates of all the atomic constituents.
In this case, the parameter $\lambda$ prescribes anchoring position of the pulling apparatus with the molecular system.
It is apparent that the consistency criterion, Eq.~(\ref{criterion2}) is satisfied because there is neither a dissipation nor a phase-volume contraction.
Accordingly, we predict that the JE, Eq.~(\ref{JE}) must work straightly in such an experimental set-up,
whereas the CWFT, Eq.(\ref{Crooks}) requires the additional symmetry condition, Eq.~(\ref{trev2}) to guarantee its applicability.

\subsection{Closed systems with thermostatted damping}
\label{Closed}

In typical experiments, the system under investigation remains immersed in a heat reservoir that energy dissipation is allowed with the surroundings.
Accordingly, apart from the manipulating force ${\cal G}^{ex}$, a dissipative mechanism ${\bf{\cal D}}$
must be taken into account in the equations of motion, Eqs.~(\ref{extHam1}) and (\ref{extHam2}),
\begin{eqnarray*}
\dot q_i &=& \frac{1}{m} p_i, \\
\dot p_i &=& -\frac{\partial V}{\partial q_i} + {\cal D}_i + {\cal G}^{ex}_i,
\end{eqnarray*}
where, for simplicity, we have set ${\cal V}^{ex}=0$.
Here, we consider that the dissipation is described by a phenomenological linear force as
\[ {\cal D}_i = - \gamma p_i \]
where the coefficient $\gamma$ is assumed to be dynamically constant but to be dependent on the reservoir temperature, $\gamma=\gamma(\beta)$.
Since the external force ${\cal G}^{ex}$ preserves the phase-volume of the system in carrying out the control work, Eq.~(\ref{LambdaC}), the consistency criterion, Eq.~(\ref{criterion2}),
is reduced to requiring
\begin{equation}\label{ex2cond1}
\int_0^\tau dt \left( \Lambda_{\cal D} + \beta \Theta \right) \equiv 0.
\end{equation}
One can calculate the Rayleigh dissipation function $\Theta$, Eq.~(\ref{Rayleigh}) to become
\[ \Theta =  2\gamma K\]
where $K=\sum p^2/2m$ is the kinetic energy of the system.
Also, the compressibility factor $\Lambda_{\cal D}$ from the damping force is given by \[
\Lambda_{\cal D} = \sum \frac{\partial}{\partial p_i}{\cal D}_i = - \gamma f \]
where $f$ denotes the degrees of freedom in the system.
Then, we find that Eq.~(\ref{ex2cond1}) brings about the NEQ energy equipartition relation,
\begin{equation}\label{equipart}
\frac{1}{\tau}\int_0^\tau dt K \equiv f\frac{1}{2}\beta^{-1}
\end{equation}
which suggests that the time-averaged kinetic-energy, $\langle K\rangle_\tau$, over the work period $\tau$ be equal to the thermal energy stored in the degrees of freedom $f$.
The condition Eq.~(\ref{equipart}) becomes an identity in the equilibrium limit, $\tau\rightarrow \infty$, but
is a strong requirement for the unidirectional work theorem over the finite period.

The applicability of the two-way work FT is further limited by the symmetry requirement, Eqs.~(\ref{symcond-exF}) and (\ref{mesorev1}).
Equation~(\ref{mesorev1}) is not satisfied in the present model because the damping coefficient  $\gamma$ is assumed to be independent of momentum.
Consequently, our theory predicts that the thermodynamic work cannot be performed reversibly even if the time-dependent force ${\cal G}^{ex}$ is symmetric in time-inversion.

The same consistency criterion, Eq.~(\ref{ex2cond1}) must be met in the nonautonomous description.
Consequently, the one-way JE would be effective in experiments where the condition of NEQ equipartition,
Eq.~(\ref{equipart}) is enforced to be satisfied.
For a bidirectional set-up the symmetric CWFT is not promising because Eq.~(\ref{mesorev1}) is not satisfied, which
is one of the symmetry conditions to be met.
The other condition from Eq.~(\ref{trev2}) does not affect the conclusion, of which explicit representation is not given in the present example.

\subsection{Computational algorithms of NEQ dynamics}
\label{SLLOD}

Here, we examine NEQ molecular dynamics (MD) of a planar Couette flow as a next, concrete example.
The fluid is assumed to be confined in spatial $y$ direction, subject to
an external shear rate $\eta$ along $x$, which is switched on, say, at $t=0$.
In the steady-state the flow velocity of the system is specified by the linear profile,
\[u_x({\bf r}) = \eta y,\]
while other spatial components are zero, where ${\bf r}$ is the field point.
In the actual simulations a difficulty arises that the shearing work generates heat in the system, which is
technically compensated by imposing a fictitious thermostat \cite{EM2008}.
We consider here the Gaussian, iso-kinetic thermostat condition that the kinetic energy in the co-moving frame with
the flow is held as constant,
\begin{equation}\label{thermostat}
\sum_{\alpha,j} \frac{p_{\alpha j}^2}{2m} \equiv \frac{1}{2}f\beta^{-1}
\end{equation}
where $p_{\alpha j}/m$ is the peculiar velocity of $\alpha$-th particle along spatial $j$ direction, $j=x,y,z$, given as
\[p_{\alpha j}/m = {\dot q}_{\alpha j}- u_j({\bf r}={\bf q}_\alpha).\]

Then, the equations of motion must be modified to incorporate both the shearing and thermostat forces.
To this end, we adopt the frequently used, SLLOD equations of motion \cite{EM2008},
\begin{eqnarray}
\dot q_{\alpha j} &=& \frac{1}{m} p_{\alpha j} + \sum_k q_{\alpha k}\frac{\partial u_j}{\partial q_{\alpha k}} \label{Gauss1}\\
\dot p_{\alpha j} &=& -\frac{\partial V}{\partial q_{\alpha j}} - \sum_k p_{\alpha k}\frac{\partial u_j}{\partial p_{\alpha k}} - \gamma p_{\alpha j}. \label{Gauss2}
\end{eqnarray}
where summation index $k$ runs over the spatial degree of freedom, $x,y,z$.
The second terms on the RHSs of Eqs.~(\ref{Gauss1}) and (\ref{Gauss2}) describe the coupling of the system to the shear field.
The third term on the RHS of Eq.~(\ref{Gauss2}) takes care of the fictitious, damping force associated with the thermostat constraint, Eq.~(\ref{thermostat}).

The frictional coefficient $\gamma$ can be specified by differentiating the iso-kinetic constraint,
Eq.~(\ref{thermostat}) with respect to time and by inserting $\dot p_{\alpha j}$ given in Eq.~(\ref{Gauss2})
into the outcome.
The result is given by
\begin{equation}\label{pdepgam}
\gamma(p;\beta) = \frac{\beta}{mf} \sum_{\alpha,j}\left\{ p_{\alpha j}\left(-\frac{\partial V}{\partial q_{\alpha j}}\right)
- \frac{1}{3} \eta p_{\alpha x} p_{\alpha y} \right\}
\end{equation}
which is evidently momentum-dependent and also temperature-dependent.
By matching the SLLOD algorithm with the extended Hamilton equations of motion, Eqs.~(\ref{extHam1}) and
(\ref{extHam2}), one can identify the non-potential terms as
\begin{eqnarray*} {\cal V}^{ex}_{\alpha j} &=& \sum_k q_{\alpha k}\frac{\partial u_j}{\partial q_{\alpha k}}  \rightarrow \eta y_\alpha\delta_{jx} = u_j({\bf q}_\alpha), \\
{\cal G}_{\alpha j}^{ex} &=&  - \sum_k p_{\alpha k}\frac{\partial u_j}{\partial q_{\alpha k}} \rightarrow -\eta p_{\alpha y} \delta_{jx}, \\
{\cal D}_{\alpha j} &=& - \gamma p_{\alpha j},
\end{eqnarray*}
where $y_\alpha$ is the $y$-coordinate of $\alpha$-th particle, $y_\alpha=q_{\alpha y}$ and $\delta_{jk}$ is the Kronecker delta.
Then, from Eq.~(\ref{control}) the work performed during the period $\tau$ by the external, shear field $\eta$
on the system is represented as
\begin{equation}\label{shearwork}
W = - \eta \int_0^\tau dt \sum_\alpha \left\{ \frac{1}{m}p_{\alpha x} p_{\alpha y} - \frac{\partial V}
{\partial q_{\alpha x}}y_\alpha \right\}
\end{equation}
where the integrand is essentially the instantaneous pressure-tensor including
the potential contribution.
The preceding work $W$ is the control work, associated with the shearing, that enters into the work-energy theorems, Eqs.~(\ref{work-energy}) and (\ref{Bochkov}).

To test how the Bochkov-Kuzovlev relation Eq.~(\ref{Bochkov}) may be fulfilled in the current system-centric picture we must inspect the validity criterion, Eq.~(\ref{criterion2}).
There appear three non-Hamiltonian sources which contribute to the compressibility factor $\Lambda$ in Eq.~(\ref{gencompr}).
It is a simple matter to calculate that the shearing fields do not affect phase volume,
\begin{eqnarray*}
&&\Lambda_{\cal G} = \sum_{\alpha,j}\frac{\partial}{\partial p_{\alpha j}}(-\eta p_{\alpha y}\delta_{jx}) = 0, \\
&&\Lambda_{\cal V} = \sum_{\alpha,j}\frac{\partial}{\partial q_{\alpha j}}(\eta y_{\alpha} \delta_{jx}) = 0.
\end{eqnarray*}
The preceding outcome manifests an interesting case that a momentum-dependent force does not give rise to phase-space contraction.
The remaining contribution in Eq.~(\ref{criterion2}) is from the thermostat force ${\cal D}$ to evaluate
\[
\int_0^\tau dt \left( \Lambda_{\cal D} + \beta \Theta \right) =0.
\]
The above expression resembles Eq.~(\ref{ex2cond1}), however, the damping coefficient $\gamma$ in ${\cal D}$ is now momentum-dependent via Eq.~(\ref{pdepgam}) and the external velocity field ${\cal V}^{ex}$ must be also taken into account in evaluating the Rayleigh dissipation function $\Theta$, defined in Eq.~(\ref{Rayleigh}).
The subsequent analysis unfolds that the compressibility factor caused by ${\cal D}$ is given by
\begin{equation}\label{LambdaD}
\Lambda_{\cal D} = -f\gamma - \sum_{\alpha j} p_{\alpha j}\frac{\partial \gamma}
{\partial p_{\alpha j}}
\end{equation}
and that $\Theta$ is calculated to be
\begin{equation}\label{ThetaD}
\Theta = \beta^{-1}\gamma f + \gamma\eta\sum_\alpha p_{\alpha x}y_\alpha.
\end{equation}
When the above results for $\Lambda_{\cal D}$ and $\Theta$ are substituted into the preceding condition, it follows that
\begin{widetext}
\begin{equation}
\int_0^\tau dt \left( \Lambda_{\cal D} + \beta \Theta \right)
= \int_0^\tau dt \left( - \sum_{\alpha j} p_{\alpha j}\frac{\partial \gamma}{\partial p_{\alpha j}} + \beta\gamma\eta\sum_\alpha p_{\alpha x}y_\alpha \right).
\end{equation}
\end{widetext}
The outcome predicts that in order for the one-way NEQ work theorem to be operative the remaining contribution from the momentum-dependence of $\gamma$ and the dissipation caused by shearing $\eta$ must sum up to vanish identically.
In addition, even if the one-way theorem holds approximately, the symmetric work FT is not likely so because the thermostatted shear flow does not possess the mesoscopic reversibility, Eq.~(\ref{mesorev1}).
The effective damping coefficient, Eq.~(\ref{pdepgam}) does not possess a definite parity under momentum-inversion.

The situation is similar in other thermostat conditions.
For instance, when the mechanical energy $H$ is fixed (i.e. iso-energetic thermostat) instead of the kinetic energy
$K$ of the system in NEQ MD simulation,
one can show that the thermostat coefficient still depends on momentum as followings, \begin{equation}
\gamma(p) = \frac{\dot W(q,p)}{2K(p)}
\end{equation}
where ${\dot W}$ is the time-rate of the control work, specified in the current case as the integrand
in Eq.~(\ref{shearwork}).

We have discussed the problem only in the system-centric picture because the coupling of the shear field to the SLLOD equations is not derivable from a nonautonomous Hamiltonian.
Thermostatted MD may provide a highly efficient and useful test bed of the NEQ equalities in deterministic many-body dynamics.

\section{Summary and conclusion}
\label{Conclusion}
We have formulated the statistical work-energy theorems for a finite system immersed in a single heat reservoir, under external manipulation, without appealing to a system--specific model.
The major drawings from our study are summarized here.

We have proposed the extended dynamics to prescribe deterministic dynamics of a thermally open system under the general NEQ conditions.
The prescribed mesoscopic dynamics embraces the coordinate-dependent (conservative), momentum-dependent (dissipative) forces, and the coupling to external time-dependent (control) fields.
The dissipative force represents the statistical correlation of the system with the surrounding reservoir and thus plays the role of a thermostat in our formulation.

We have endeavored to formulate the extended dynamics both in the system-centric picture and in the nonautonomous Hamiltonian picture.
In the former the mechanical energy of the system is an instant value of the bare Hamiltonian, excluding the time-dependent perturbation, of which dynamical variables obey the extended Hamilton equations of motion [Eqs.~(\ref{extHam1}) and (\ref{extHam2})].
The energy is not conserved but complies with the mechanical work-energy theorem [Eq.~(\ref{work-energy})].
The work-energy balance was also formulated alternatively in the avenue of the nonautonomous Hamiltonian [Eq.~(\ref{nonautoW-E})].
The resulting parametric work is represented as the negative integral of the coordinates over the variation of the external forces [Eq.~(\ref{defwork2})], differently from the system-centric definition of work as the integral of the forces over the displacements of coordinates [Eq.~(\ref{defwork1})].

The mechanical work-energy theorems are exact but they merely serve as a theoretical guidance due to
the enormous degrees of freedom in the finite system (due to fluctuation in small systems).
In order to account for the NEQ work measurement of the system controlled by the external perturbation a statistical description must take over.
We have performed the statistical average of the mechanical work-energy theorems, adopting Jarzynski's mathematical  recipe of the exponential work, over the NEQ distribution of the identically prepared ensemble of the finite system.
The NEQ phase-space density obeys the generalized Liouville dynamics, of which generator of convective time-development turns out to be the non-vanishing compressibility factor.
Consequently, we have derived the Bochkov-Kuzovlev equality [Eq.~(\ref{Bochkov})] in the system-centric description and the JE [Eq.~(\ref{JE})] in the nonautonomous description.
Our formulation is highlighted by the physical criterion [Eq.~(\ref{criterion2})], being consistent with the detailed balance condition [Eq.~(\ref{criterion3})] for the generalized Liouville dynamics [Eq.~{\ref{GLiouville})], of which satisfaction assures the NEQ equalities as rigorous theorems in a thermally open system.

The momentum-dependent, damping force renders the associated NEQ thermodynamic process typically irreversible.
Nevertheless, the extended equations of motion may be still symmetric under time reversal, conditioned on that
the damping force is an even function of momentum and also that the other external fields are symmetric under
time inversion.
Such a constrictive symmetry is archived in the generalized Liouville dynamics if the compressibility factor
changes its sign under time reversal [Eq.~(\ref{trevcomp})].
Then, the time-reversed ensemble density can propagate backward in time under the same propagator.
In such a system of possessing the mesoscopic reversibility, we have shown that a NEQ work measurement may be
performed in the bidirectional manner with conforming to either the Bochkov-Kuzovlev WFT [Eq.~(\ref{BKWFT})] or
the CWFT [Eq.~(\ref{Crooks})], depending on choice of the picture.                                                The symmetric CWFT yields the JE as a corollary, however, unlike the usual interpretation we note that the latter
is more general in the sense that it can be applied to a time-asymmetric transformation.

In conclusion, we have explored the NEQ work theorems by directly taking the NEQ ensemble average of
the mechanical energy balances.
Consequently, a consistency condition has been derived which tightens the Bochkov-Kuzovlev-Jarzynski-Crooks NEQ equalities to be legitimate in thermally open, finite systems.
The condition affirms that the unidirectional work theorems for irreversible transformation are contingent on that the contracted phase-volume from all involved nonconservative forces must be precisely offset by the dissipated power scaled by the equilibrium temperature over the work period, constituting the detailed balance in thermostatted, deterministic dynamics.
The criterion is also implemented in the bidirectional work FTs, however, with the additional symmetry requirement that the dynamics of the system be invariant under time reversal even in the presence of the dissipation and nonpotential manipulating forces.
We hope that our unveiling provides researchers with a useful, theoretical appraisal of
the NEQ work theorems in real or computer experiments.

\begin{acknowledgments}
The author is grateful to Gary P. Morriss for providing a useful discussion, in particular about
thermostatted NEQ dynamics.
\end{acknowledgments}

\section*{References}

\end{document}